\documentclass[aps,amssymb,pra,
showpacs,twocolumn]{revtex4} 

\usepackage{graphicx}
\begin{document}

\title{Quantitative analysis of directional spontaneous emission
spectra from light sources in photonic crystals}

\author{Ivan S. Nikolaev}
\email{i.nikolaev@amolf.nl}

\author{Peter Lodahl}
\author{Willem L. Vos}

\affiliation{Complex Photonic Systems (COPS), Department of
Science and Technology, and MESA$^+$ Institute of Nanotechnology,
University of Twente, PO Box 217, 7500 AE Enschede, The
Netherlands.} \homepage{www.photonicbandgaps.com}

\altaddress[Present address: ]{FOM Institute for Atomic and
Molecular Physics (AMOLF), P.O. Box 41883, 1009 DB Amsterdam, The
Netherlands}

\begin{abstract}
We have performed angle-resolved measurements of spontaneous-emission
spectra from laser dyes and quantum dots in opal and inverse opal
photonic crystals. Pronounced directional dependencies of the
emission spectra are observed: angular ranges of strongly reduced
emission adjoin with angular ranges of enhanced emission. It appears
that emission from embedded light sources is affected both by the
periodicity and by the structural imperfections of the crystals: the
photons are Bragg diffracted by lattice planes and scattered by
unavoidable structural disorder. Using a model comprising diffuse
light transport and photonic band structure, we quantitatively
explain the directional emission spectra. This work provides detailed
understanding of the transport of spontaneously emitted light in real
photonic crystals, which is essential in the interpretation of
quantum-optics in photonic band-gap crystals and for applications
wherein directional emission and total emission power are controlled.
\end{abstract}


\maketitle

\section{Introduction}
Photonic crystals attract much attention both in academia and in
industry because they offer exciting ways of manipulating photons
\cite{Soukoulis01}. Periodic variations of the refractive index in
photonic crystals on a length scale of the wavelength of light cause
optical Bragg diffraction and organize the photon dispersion relation
in bands, analogous to electron bands in semiconductors
\cite{Bykov75,Yab87,John87}. Frequency windows, called stop bands,
appear in which there are no modes for certain propagation
directions. Photonic structures already serve as a base for
controlling the propagation of light. Of even greater interest are
three-dimensional (3D) photonic crystals that summon novel
opportunities in the case of a photonic band gap (PBG) - a frequency
range where no modes exist at all. The existence of a gap in the
density of photonic modes leads to novel quantum-optical phenomena
such as spontaneous emission inhibition and light localization
\cite{Bykov75,Yab87,John87,Vats02,Yang03}.

Control over the radiative decay rate of spontaneous emission is
of keen interest for applications, and therefore emission
properties of sources such as atoms, dyes, and quantum dots are
intensively investigated. According to Fermi's `Golden Rule', this
decay rate is proportional to the local radiative density of
states (LDOS) that counts the number of electromagnetic modes
wherein photons can be emitted at a specific location of the
source. In 3D photonic crystals pronounced variations of the LDOS
with frequency are predicted even in the absence of a PBG
\cite{Sprik96,Busch98}, which give rise to angle-independent
variations of the total emission. Recently LDOS effects on
spontaneous emission have been experimentally demonstrated: Using
inverse opals photonic crystals considerable variations of the
emission rates in large band widths were obtained in both
continuous-wave (cw) total-emission power experiments
\cite{Koenderink02} and in time-resolved lifetime experiments
\cite{Lodahl04}. While lifetime experiments provide a direct
measurement of decay rates, it is important to quantitatively
interpret concomitant emission spectra, for instance, to confirm
that light sources inside the crystal are probed. Cw experiments,
on the other hand, rely on a comparison of angle-integrated
spectra with a homogeneous medium. In the latter case a complete
understanding of all angle-dependent effects, that is, Bragg
diffraction on the propagation of light is crucial. In this paper
such a quantitative analysis is presented.

In contrast to decay rates, emission spectra of sources embedded in
photonic crystals are strongly directional
\cite{Bog97,Blanco98,Megens99,Vlasov00,Schriemer01,Lin02}. Particular
frequency ranges of the spectra are suppressed in certain directions
forming stop bands, whose center frequencies and the widths are
described by the photon dispersion relation. Besides Bragg
diffraction, which is an effect of the order of the periodic
structure, light propagating inside the structure also \emph{feels}
disorder: polydispersity, roughness and misarrangements of the
building blocks~\cite{Koenderink04}. This unavoidable disorder
affects the interference-induced properties of photonic crystals.
Previous work on the effect of disorder on spontaneous emission
includes the realization that disorder determines the depth of stop
bands~\cite{Megens99,Schriemer01}. Furthermore, the first
observations of enhanced emission in the range of first and
second-order stop bands were also related to disorder-induced
redistribution of emitted photons \cite{Bechger04}. While the
propagation of light from external light sources has been studied in
great detail \cite{Koenderink02b,Galisteo03,Koenderink03}, a
quantitative explanation of the behavior of light emission from
internal sources has lacked sofar.

Here we present strongly frequency-dependent angular distributions of
spontaneous emission from a laser dye in polystyrene opals and from
quantum dots in titania inverse opals in the frequency range around
the first-order Bragg diffraction (L-gap). We compare the data to a
theoretical model that unifies effects of structural disorder and
photonic crystal properties \cite{Koenderink03}. Angle-dependent
internal reflection due to the photonic gaps plays a key role in our
model. The theory quantitatively explains both the enhancement and
the reduction of light along certain propagation directions that were
observed experimentally. The excellent agreement confirms that the
propagation of light in a photonic crystal is well understood for
frequencies around the L-gap. Furthermore, we show that by analyzing
the exit emission distributions, one can reveal stop bands in the
quantum dot spectra. Such an analysis should be carefully performed
before any quantum-optical experiments since it unambiguously proves
the effect of the photonic crystal on emission. We finally discuss
the applicability of photonic crystals for improvement of the
emission efficiency of light sources.

\section{Experimental details}
\subsection{Samples}

We have studied emission from dyes in polystyrene opals and from
quantum dots in titania inverse opals. The polystyrene opals are fcc
crystals of close-packed polystyrene spheres prepared from a
colloidal suspension by self-assembly. The titania inverse opals are
fcc structures of close-packed air spheres in a solid matrix of
TiO$_2$. Details of the preparation and characterization of the opals
and inverse opals can be found in Ref. \cite{Wijnhoven01}. We have
studied $4$ polystyrene samples with lattice parameters $\emph{a}$ =
178 $\pm$ 3 and 365 $\pm$ 5 nm and 8 titania samples with lattice
parameters $\emph{a}$ = 370, 420, 500, 580, and 650 $\pm$ 10 nm. All
samples have typical dimensions of 2x2x0.2 mm$^3$ and contain
high-quality domains with diameters larger than 50 $\mu$m. These
domains have flat faces normal to the 111 crystal axis, which is
evident from SEM images. The other crystalline axes are randomly
oriented.

The polystyrene opals were doped with the laser dye Rhodamine 6G
(R6G) by soaking them for 30 minutes in a dilute solution $(10^{-6}$
mol/l) of the dye in ethanol. Afterwards the samples were rinsed in
ethanol and dried to remove the dye from the sample surface. To
estimate an upper bound to the density of the dye, we assume that the
infiltrating solution completely fills the air voids in the opals,
and that, in the process of drying, the dye molecules distribute
uniformly on the surfaces of the spheres inside the opals. Knowing
the volume of the air voids in each unit cell of the opals
($0.26~a^3$) and the dye concentration, we arrive at no more than 10
dye molecules per unit-cell inner surface \cite{inner-area} (for
opals with lattice parameter \emph{a} = 365 nm). While this surface
density increases linearly with the lattice parameter $a$, the
average distance between the dye molecules remains more than 2 orders
of magnitude away from the typical intermolecular distances where
reabsorption and energy-transfer processes could play a role
\cite{Lakowicz}. Before emission experiments, we performed a
selective bleaching \cite{bleaching} of the dye at the surface of
opal photonic crystals in order to ensure that the emission is
recorded only from the bulk of the crystal, and not from the crystal
surface \cite{Megens99}. The whole surface was illuminated for up to
1 hour by an intense laser beam at the Bragg angle for the pump
frequency. At this angle the pump intensity decreases exponentially
with depth, which implies that the dye bleaches within the first few
crystal layers. However, our experimental results were found to be
almost independent of the amount of bleaching, thus indicating that
emitters in the bulk of the photonic crystals provide the dominating
contribution to the measured emission intensity.

The titania inverse opals were infiltrated for 24 hours with a
colloidal suspension of ZnSe-coated CdSe quantum
dots~\cite{Dabbousi97,Donega03} in a mixture of $50\%$ chloroform and
$50\%$ butanol. Afterwards the samples were rinsed in chloroform and
dried. The concentration of the infiltrating solution was $10^{-7}$
mol/l. We used the same estimation in order to get the maximum number
of quantum dots per unit cell in the inverse opals. For samples with
the largest lattice parameter $a$ = 650 nm, this concentration is 15
quantum dots per unit cell and is sufficiently low to avoid
energy-transfer processes and reabsorption. In order to minimize
oxidation and contamination of the quantum dots, the inverse opals
were infiltrated with the quantum dots in a nitrogen-purged glove box
and held in a sealed chamber under a 1.7 mbar nitrogen atmosphere
during the optical measurements. No bleaching at the surface layers
could be performed for the quantum dots, however a detailed analysis
of the angular emission data (see below) demonstrates that light
emission from the quantum dots from the bulk dominates.

\begin{figure}
\includegraphics[width=0.25\textwidth]{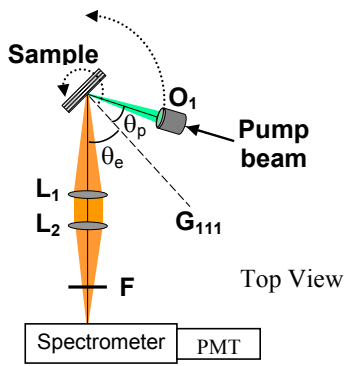}
\caption{(color online) \label{fig:set-up} Scheme of the
experimental set-up. The pump beam is focused onto the sample at
incident angle $\theta_p$ by the objective $O_1$ (f = $7.3$ cm, NA
= 0.05). Luminescence within a cone centered at detection angle
$\theta_e$ relative to the surface normal is collected by the lens
$L_1$ (f = 12 cm) and imaged on the spectrometer slit by the lens
$L_2$ (f = 12 cm). A colour filter $F$ prevents scattered pump
light from entering the prism spectrometer. The angle $\theta_e$
is varied by rotating the rotation stage, which carries the sample
holder, the fiber, and the objective $O_1$, whereas the incident
angle $\theta_p$ is kept fixed.}
\end{figure}

\subsection{Experimental set-up}

Figure~\ref{fig:set-up} shows the experimental set-up used to measure
emission from light sources inside photonic crystals. The sources
inside the crystal are excited by a cw Ar-ion laser ($\lambda =
497$~nm) with the power at the sample around 10 $\mu$W. At this pump
power, we do not observe any effects of bleaching of the dye during
the emission experiments. To focus the pump beam on the sample
surface a fiber-coupled microscope objective is used. The beam is
focused to a spot of about 30 $\mu$m in diameter at an incident angle
$\theta_p$ relative to the surface normal, usually
$\theta_p\approx~25^0$. In order to acquire emission spectra as a
function of the detection angle $\theta_e$ relative to the surface
normal, the sample is mounted on a rotation stage. The surface normal
is parallel to the 111 reciprocal lattice vector G$_{111}$. In order
to illuminate the same area irrespective of $\theta_e$, the
fiber-coupled objective ($O_1$) is mounted on the same rotation stage
as the sample. In this way, the angle of incidence $\theta_p$ between
the pump beam and the surface normal is kept constant. The advantage
over previous experiments, where the sample was rotated with respect
to both the pump and detection beams, is that the absolute intensity
of the angle-dependent spectra can be reliably compared. The position
of the pump spot on the sample is monitored with a microscope. The
emitted light is collected within a cone of $15^0$ full width around
the angle $\theta_e$, imaged on the slit of the spectrometer with 4
nm resolution, and detected by the photomultiplier tube (PMT). The
angle-resolved spectra are usually measured at the detection angles
from $0^0$ to $75^0$ at intervals of $15^0$. The measured spectra are
corrected for the dark count of the PMT. The shapes of the spectra
are confirmed to be independent of the pump intensity, and the
emitted intensity is linear with the pump power.

\section{Diffuse light transport in photonic crystals}
\subsection{Escape function}
In real photonic structures, defects in the arrangement of the
building blocks are always present and cause random multiple
scattering of light. This means that all light emitted in such
photonic structures becomes diffuse on length scales equal to the
transport mean free path \emph{l}, which is often much smaller than
the thickness of the sample \emph{L}. For example, our opals and
inverse opals have mean free paths of about 15 $\mu$m
\cite{Koenderink00}, whereas the thickness of the samples is about
200 $\mu$m. Thus, even though photons generated inside a photonic
crystal are diffracted by the crystal structure, this effect is
smeared out by the random multiple scattering while the photons
propagate through the bulk towards the crystal surface. Only at
distances from the surface \emph{z} smaller than \emph{l}, where the
photons emanate ballistically towards the crystal-air interface after
a last scattering event, the effect of Bragg diffraction is not
destroyed by the scattering. Hence the diffuse emission acquires a
directional dependence only when it exits the
crystal~\cite{Koenderink03}.

We consider the ratio of the mean free path $l$ to the attenuation
length for Bragg diffraction $L_B$ in order to estimate the
attenuation of emission caused by Bragg diffraction, as proposed in
Ref. \cite{Schriemer01}. Since the mean free path \textit{l} is
larger than the Bragg attenuation length $L_B$ (typically {$l / L_B
\sim 2-5$} \cite{Koenderink00}), an attenuation in the stop band
equal to $1- L_B / l$ = $50\%$ to $80\%$ is predicted, which is in
agreement with our observations. As will be discussed later, the
stop-band attenuations are obtained directly from reflectivity
measurements, therefore, the mean free path $l$ is not an explicit
parameter in our theoretical model.

In the present work we investigate directional properties of light
emitted by sources from 3D photonic crystals and compare to a
model of diffuse light transmission through opaque
media~\cite{Lag89,Zhu91,Durian94} extended to photonic
crystals~\cite{Koenderink03}. Based on the diffusion theory, the
intensity of light $I(\omega, \mu_e)$ with frequency $\omega$ that
exits the sample at external angles between $\theta_e =
\cos^{-1}(\mu_e)$ and $\cos^{-1}(\mu_e + d\mu_e)$ relative to the
surface normal is equal to
\begin{equation}I(\omega,\mu_e)d\mu_e = I_{tot}(\omega)P(\omega,\mu_e)d\mu_e.
\end{equation}
Here $I_{tot}(\omega)$ is the total spontaneous emission power
that is the spectrum of the light sources integrated over the exit
angles $\theta_e$. For sources with a low quantum efficiency or
with inhomogeneously broadened spectra, $I_{tot}(\omega)$ is
proportional to the LDOS~\cite{Koenderink02}. The distribution
$P(\omega,\mu_e)$ is defined as
\begin{equation}\label{escapefunction}
P(\omega,\mu_e)=\mu_e\frac{n^2_e}{n^2_i}\biggl(\frac{1+\bar{R}(\omega)}{1-\bar{R}(\omega)}+\frac{3}{2}\mu_i\biggr)
[1-R(\omega,\mu_i)],
\end{equation}
where $n_e$ and $n_i$ are average refractive indices outside and
inside the sample \cite{average refr index}, respectively. $\mu_e$
and $\mu_i$ are related by Snel's law. $R(\omega, \mu_i)$ is an
angle-dependent internal-reflection coefficient that yields an
angle-averaged internal-reflection coefficient $\bar{R}(\omega)$:
\begin{eqnarray}\bar{R}(\omega)=\frac{3C_2(\omega)+2C_1(\omega)}{3C_2(\omega)-2C_1(\omega)+2}~,
\\C_n(\omega)=\int^1_0{R(\omega, \mu_i)\mu_i^nd\mu_i}~.
\end{eqnarray}
From the diffusion theory, $\bar{R}(\omega)$ determines the so-called
extrapolation length that sets the boundary conditions of the diffuse
intensity~\cite{Lag89,Zhu91,Durian94}. The normalized function
$P(\omega, \mu_e)$ describes the distribution of emission intensity
over the escape angles and will be called the `escape function'. In
absence of reflection effects, the escape distibution tends to the
well-known Lambertian distribution of diffuse surfaces.

In random media such as powders or macroporous sponges the
internal-reflection coefficient $R(\omega, \mu_i)$ is barely
frequency dependent \cite{Schuurmans99}, and propagation through the
interface is well described by Fresnel reflection model assuming an
average refractive index. The angular dependence of the escape
function $P(\omega, \mu_e)$ agrees well with experiments on random
media \cite{Zhu91,Durian94}. For highly dispersive photonic crystals,
however, Fresnel model cannot describe the internal reflection since
light escaping from a depth $z < l$ from the crystal surface is Bragg
attenuated for angles and frequencies inside a stop band. We model
the strong angle and frequency dependent internal reflection with
photonic band structures. At a particular frequency $\omega^*$ where
a stop band is present, the internal reflectivity $R(\omega^*,
\mu_i)$ blocks the emission in the directions of the stop band (a
range of $\mu_i^*$'s related to $\omega^*$ by the photonic band
structure) and therefore reduces the escape function $P(\omega^*,
\mu_e^*)$, \emph{cf.} Eq. (\ref{escapefunction}). The presence of the
stop band raises the angle-integrated reflectivity
$\bar{R}(\omega^*)$, which enhances the escape function $P(\omega^*,
\mu_e)$ for angles outside the stop band. Thus, the escape function
is strongly non-Lambertian in a photonic crystal, showing clear
suppressions or enhancements.

\begin{figure*}
\includegraphics[width=0.7\textwidth]{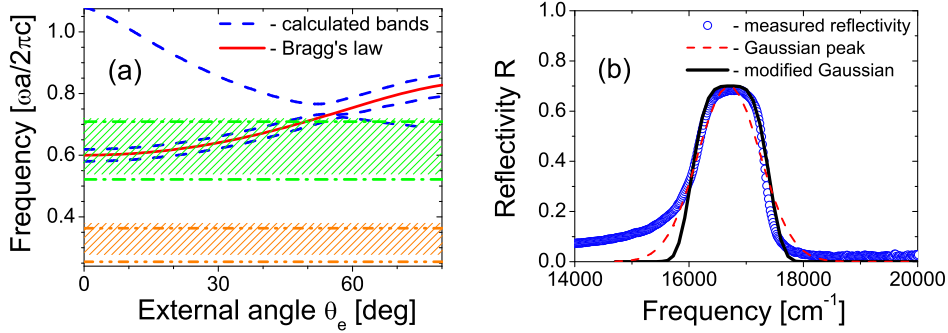}
\caption{(color online) \label{fig:reflectivity, bands} (a)
Photonic band structure for polystyrene opals (blue dashed
curves), center frequency of the stop band vs. detection angle
$\theta_e$ according to Bragg's law (red curve), the frequency
range of R6G emission is shown for opals with lattice parameters
\emph{a} = $178$ and $365$ nm (hatched regions between the orange
dash-dot-dotted lines and the green dash-dotted lines,
respectively). (b) Normal-incidence reflectivity as a function of
frequency for polystyrene opals with a lattice parameter \emph{a}
= $365$ nm and refractive-index contrast m = $1.59$. Blue circles
are measured values, the black curve is a fit with the modified
Gaussian reflectivity peak (Eq. \ref{modGauss1}), the red dashed
curve is a fit with a Gaussian reflectivity peak.}
\end{figure*}

\subsection{Internal-reflection coefficient}
In order to model the internal-reflection coefficient $R(\omega,
\mu_i)$, we have taken into consideration calculated photonic band
structures. Figure~\ref{fig:reflectivity, bands}(a) shows the
photonic band structure for polystyrene opals calculated along the LU
and LK lines in reciprocal space in the frequency range around the
first-order stop gap (L-gap) that is due to optical Bragg diffraction
by (111) planes parallel to the sample surface. The emission spectrum
of R6G is in the low-frequency limit relative to the stop bands of
opals with a lattice parameter $\emph{a}$ = 178 nm (region confined
by the orange dash-dot-dotted lines). Consequently, this sample is
effectively homogeneous for the emission frequencies, and therefore
it can serve as a reliable reference. For these nonphotonic crystals
we used Fresnel model in order to describe the internal reflection.
In Figure~\ref{fig:reflectivity, bands}(a) one can also see the
frequency gap between the two lowest bands (blue dashed curves),
which obeys Bragg's law (red curve) within the frequency range of R6G
emission for the opals with a lattice parameter $\emph{a}$ = 365 nm
(region confined by the green dash-dotted lines). Therefore, the
angular dependence of the center frequency of the L-gap is modelled
with the red curve, i.e.: $\omega_C(\mu_i) =
\omega_C(\mu_i=1)/\mu_i$. To investigate the frequency dependence of
the reflectivity, we have performed reflectivity experiments on the
samples using external incident plane waves, since this technique
reveals the center frequencies and the widths of stop bands
\cite{Thijssen99,Galisteo03}. Figure~\ref{fig:reflectivity, bands}(b)
shows a normal-incidence reflectivity spectrum measured from an opal
with the lattice parameter $\emph{a}$ = 365 nm (blue circles). The
reflectivity peak is not fitted well with a Gaussian (red dashed
curve). As an improved model, we propose a modified Gaussian
(Fig.~\ref{fig:reflectivity, bands}(b), black curve):
\begin{equation}\label{modGauss1}
R_1(\omega,\mu_i)=A_1(\mu_i)\exp{\bigg[-\frac{(\omega-\omega_C(\mu_i))^4}{2(\Delta\omega_C(\mu_i))^4}}\bigg],
\end{equation}
where $A_1(\mu_i)$ is the magnitude of the internal-reflection
coefficient and $\Delta\omega_C(\mu_i)$ is the width parameter. This
peak shape is seen to fit the measurements well for frequencies $>$
16000 cm$^{-1}$. At frequencies below the stop band, i.e. below 16000
cm$^{-1}$ for these particular samples with \emph{a} = 365 nm, a
deviation from the model is observed. We attribute this deviation to
Fresnel reflection, which is important only in the low-frequency
limit and therefore is not relevant for the escape function of
photonic samples. The width of the L-gap $\Delta\omega_C(\mu_i)$
hardly varies with $\mu_i$ within the range of the dye emission,
therefore it is taken to be constant in our model. The magnitude of
the internal-reflection coefficient $A_1(\mu_i)$ decreases with
$\mu_i$ because at larger internal angles $\theta_i=\cos^{-1}(\mu_i)$
the path length for the light to become Bragg attenuated increases
with $\mu_i$, and this increases the probability of scattering at $z
< L_B$. The value of $A_1(\mu_i)$ at $\mu_i=1$ is taken from the
normal-incidence reflectivity experiments. Thus, we have:
$A_1(\mu_i)=A_1(\mu_i=1)\cdot\mu_i$ and $A_1(\mu_i=1)=0.7$, see
Fig.~\ref{fig:reflectivity, bands}(b).

Emitted light that is scattered within a distance $L_B < z < l$
towards the exit interface can also be redirected by Bragg
diffraction by the sets of $(11\bar{1})$ planes, which are
oriented at $\theta_i=70.5^0$ to the (111) planes and the surface
normal. The internal-reflection coefficient $R_2(\omega,\mu_i)$
for Bragg diffraction by $(11\bar{1})$ lattice planes is modelled
similarly to $R_1(\omega,\mu_i)$. Taking into account that we
measure emission from many randomly-oriented crystal domains in
azimuthal directions, the reflectivity $R_2(\omega,\theta_i)$ is
averaged over the azimuthal angles $\phi$ between the LK and LU
lines in reciprocal space, yielding:
\begin{equation}\label{modGauss2}
R_2(\omega,\theta_i)=\frac{3}{\pi}\int^{\pi/3}_0{A_2(\theta_i,\phi)\exp{\bigg[-\frac{(\omega-\omega_C(\theta_i,\phi))^4}
{2(\Delta\omega_C(\theta_i))^4}}\bigg]d\phi.}
\end{equation}
The magnitude $A_2(\theta_i,\phi)$ is modelled as
$A_2(\theta_i,\phi)=A_2(70.5^0,0^0)\cdot cos(\theta_i-70.5^0)\cdot
cos(\phi)$ with $A_2(70.5^0,0^0)=0.7$. The total internal-reflection
coefficient $R(\omega,\mu_i)$ is calculated as a sum of the
$R_1(\omega,\mu_i)$ and $R_2(\omega,cos(\theta_i-70.5^0))$ modified
Gaussian peaks. We expect this model of the angle- and
frequency-dependent internal reflectivity to capture the essential
frequency dependence of the first-order photonic stop bands in
polystyrene opals.

In the case of the titania inverse opals we apply the same escape
model to explain our experimental data. However, in calculating the
internal-reflection coefficient $R(\omega,\mu_i)$, Bragg diffraction
from other lattice planes must also be included. This difference
compared to polystyrene opals appears since titania inverse opals are
more strongly photonic and the measurements were performed at higher
reduced frequencies ($a/\lambda=\omega a/2\pi c$). Moreover, the
resulting stop bands occur at lower detection angles $\theta_e$ in
these crystals than in the polystyrene opals, as a consequence of the
lower average refractive index. Therefore Bragg's law is not a
sufficient approximation and is not used to model the angular
dependence of the stop bands in the titania inverse opals. Instead,
the full band structure model is employed, in which we take into
account multiple Bragg wave coupling \cite{Schriemer01,vDriel00}. For
the inverse opals, this model was already successfully tested on
diffuse transmission experiments \cite{Koenderink03}.

\begin{figure*}
\includegraphics[width=0.75\textwidth]{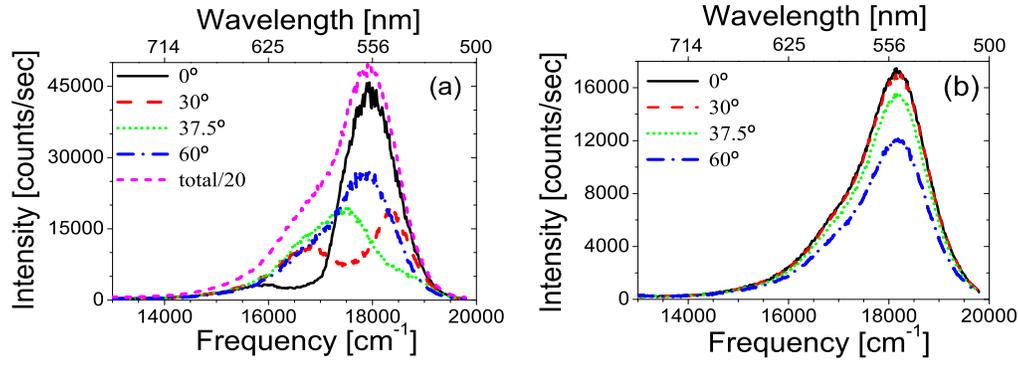}
\caption{(color online) \label{fig:raw spectra}  Emission spectra
of R6G in polystyrene opals with lattice parameters \emph{a} =
$365$ nm (a) and \emph{a} = $178$ nm (b). The black curves are
obtained at $\theta_e = 0^0$, the red dashed curves at $\theta_e =
30^0$, the green dotted curves at $\theta_e = 37.5^0$, and the
blue dash-dotted curves at $\theta_e = 60^0$. The magenta
short-dashed curve in (a) indicates the total emission spectrum
$I_{tot}(\omega)$.}
\end{figure*}

\section{Results and Discussion}
\subsection{Spontaneous emission of R6G in polystyrene opals}
Reflectivity measurements at normal incidence of polystyrene opals
(Fig.~\ref{fig:reflectivity, bands}(b)) reveal that the relative
width of the first-order stop band is $\Delta\omega/\omega$ $\approx
0.075$. For opals with a lattice parameter \emph{a} = $365$ nm this
means a stop band in the range 16100 - 17300 cm$^{-1}$ for light
escaping the crystal normally to the surface. The dye R6G emits in
the range of 15000 - 20000 cm$^{-1}$, and hence we expect to observe
directional-dependent emission of the dye from the opals with
\emph{a} = $365$ nm. Figure~\ref{fig:raw spectra}(a) displays the
emission spectra at selected detection angles for such doped opals.
It is clearly seen that the shapes of the spectra are affected by the
photonic crystal. The emission is suppressed by the crystal for
$\theta_e = 0^0$ in the spectral range from 16000 to 17500 cm$^{-1}$.
With increasing angle $\theta_e$ the low-frequency parts of the
emission recover, and the suppressed emission range shifts to higher
frequencies, as expected from Bragg's law for a photonic stop band
from a single set of lattice planes. In contrast, the shape of the
spectra from an opal with a lattice parameter \emph{a} = $178$ nm
remains unchanged (Fig.~\ref{fig:raw spectra}(b)). The sample is not
photonic for the frequency range considered: the frequencies of R6G
emission lie far below the first-order stop band in the opal with
this lattice parameter.

\begin{figure}
\includegraphics[width=0.35\textwidth]{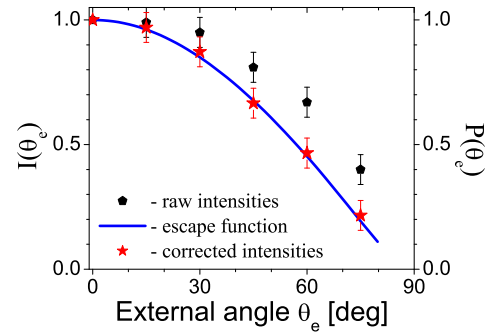}
\caption{(color online) \label{fig:det-eff}  Angular distribution
of the R6G emission from a polystyrene opal with a lattice
parameter \emph{a} = $178$ nm. The black pentagons indicate the
intensity measured at the spectral maximum ($\omega$ = 17860
cm$^{-1}$ or $\lambda$ = 560 nm), the blue curve is the calculated
escape function with Fresnel internal-reflection coefficient.
Measured intensities corrected for the detection efficiency of the
set-up are displayed as red stars. All data are normalized at
$\theta_e = 0^0$.}
\end{figure}

Before studying spontaneous emission from photonic samples we have
verified the applicability of the above-mentioned model of diffuse
light transport on the nonphotonic, reference samples. We use Fresnel
reflection to describe the angular-dependent internal-reflection
coefficient, taking an average refractive index $n_{av}=1.44$, which
is derived from the polystyrene filling fraction $\varphi\approx74\%$
in opals and the refractive index of polystyrene n = $1.59$. We
record the intensity at the maximum of the emission spectrum as a
function of the exit angle $\theta_e$ relative to the measurement at
$\theta_e = 0^0$. The relative intensity $I(\theta_e)$ is compared to
the escape function $P(\theta_e)$ in Figure~\ref{fig:det-eff}. While
the expected decrease with angle is observed, it is clear that the
calculated intensity differs systematically from the measured data.
This deviation appears to be caused by an angle-dependent detection
efficiency as a result of an increase with $\theta_e$ of the
projection of the spectrometer slit on the sample. Correcting the
measured intensity $I(\omega,\mu_e)$ for the detection efficiency
$D(\mu_e)$ (see Appendix) yields the corrected intensity
$I_c(\omega,\mu_e) = I(\omega,\mu_e)/D(\mu_e)$ displayed as red
symbols in Fig.~\ref{fig:det-eff}.  The agreement between the
corrected intensity and the calculated escape function is excellent.
With the proper account of the detection efficiency, the angular
distribution of emission escaping the reference samples is thus fully
understood. In all experimental data presented in the remainder of
this paper the detection efficiency has been included.

\begin{figure}
\includegraphics[width=0.35\textwidth]{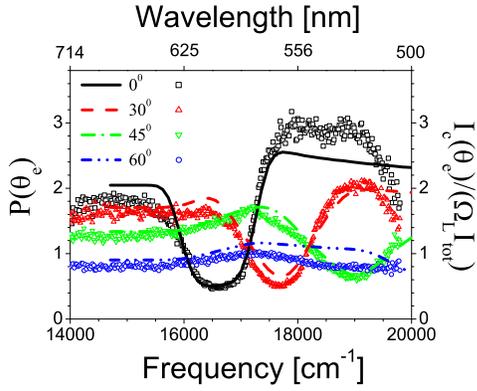}
\caption{(color online) \label{fig:spectra-ratios} R6G emission
from polystyrene opal with lattice parameter a = $365$ nm. The
scatter plots represent the measured spectra corrected for the
angular aperture of the collecting lens $\Omega_L$ and divided by
the total emission spectrum $I_{tot}(\omega)$. The calculated
escape functions are plotted with solid curves. The black solid
curve and squares are for $\theta_e = 0^0$, red dashed curve and
triangles are for $\theta_e = 30^0$, green inverted triangles and
solid curve are for $\theta_e = 45^0$ and blue circles with
dash-dotted curve are for $\theta_e = 60^0$.}
\end{figure}

In the case of the photonic samples the exit distribution of emission
strongly depends on the frequency $\omega$ as mentioned above:
$P(\omega,\theta_e)=I_c(\omega,\theta_e)/I_{tot}(\omega)$. The total
emission spectrum $I_{tot}(\omega)$ is determined by discretely
summing the angle-resolved spectra $I_c(\omega,\theta_e)$ weighted by
$2\pi\sin(\theta_e)d\theta_e$ to approximate the integration over
$2\pi$ solid angle. The spectra from Figure~\ref{fig:raw spectra}(a)
divided by the total emission spectrum $I_{tot}(\omega)$ are plotted
in Figure~\ref{fig:spectra-ratios} (symbols) together with the
calculated escape function $P(\omega,\theta_e)$ (curves). We observe
a good agreement between our experiment and theory. The escape
function hardly varies with frequency in the low-frequency region
$\leq$ 15600 cm$^{-1}$, while it still depends on the detection
direction. In contrast, at higher frequencies strong variations are
seen compared to the low-frequency range. At the exit angle $\theta_e
= 0^0$, the escape function is significantly reduced in the spectral
range from 16000 to 17500 cm$^{-1}$ by the stop band centered at
$\omega$ = 16700 cm$^{-1}$ due to internal Bragg diffraction, which
is described by the term $(1-R(\omega,\mu_i))$ in
Eq.~(\ref{escapefunction}). The change of the center frequency as
well as the decrease in the attenuation of emission inside the stop
band with increasing exit angle $\theta_e$ are well described in our
model by the frequency and angular dependent internal-reflection
coefficient $R(\omega,\mu_i)$. At $\theta_e = 60^0$, the stop band
has moved out of the spectral range of R6G.

\begin{figure}
\includegraphics[width=0.35\textwidth]{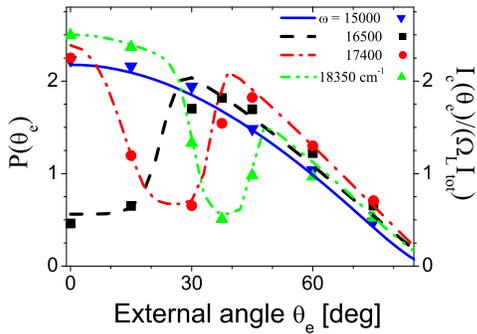}
\caption{(color online) \label{fig:escape-angular} Intensity
ratios $I_c(\theta_e)/I_{tot}$ corrected for the lens aperture
$\Omega_L$ as a function of the exit angle $\theta_e$ for a
polystyrene opal with a lattice parameter $a$ = $365$ nm for
frequencies $\omega$ = 15000, 16500, 17400 and 18350 cm$^{-1}$
(blue inverted triangles, black squares, red circles and green
triangles). The corresponding curves represent the calculated
escape distributions (for reduced frequencies $\omega a/2\pi c$ =
0.55, 0.6, 0.64 and 0.67, respectively).}
\end{figure}

Figure~\ref{fig:spectra-ratios} also shows a peculiar feature: the
frequency ranges where the emission is inhibited along certain
directions, adjoin with the ranges where emission is \emph{increased}
along the same directions. Such an increase appears at the blue side
of the stop band at $\theta_e=0^0$ and $30^0$, and at the red side of
the stop band at $\theta_e = 45^0$ and $60^0$. This enhanced escape
probability in the frequency range 16000 - 19500 cm$^{-1}$ along
directions that do not coincide with the stop band is reflected in
our model for $P(\omega,\mu_e)$ (Eq.~(\ref{escapefunction})) as an
increase of the angle-averaged internal-reflection coefficient
$\bar{R}(\omega)$. The good agreement between experiments and theory
confirms a qualitative attribution of such enhancements to diffuse
escape effects \cite{Bechger04}. Moreover it unambiguously
demonstrates that our experimental observation of the emission
enhancement is not due to Bragg standing wave effects proposed in
Ref. \cite{Galstyan00}, but is related to diffusion of light.
In Figure~\ref{fig:escape-angular} we compare the experimentally
determined intensity distributions (symbols) for several fixed
frequencies with the calculated ones (solid curves) as functions
of the exit angle $\theta_e$. The experimental values of
$P(\theta_e)$ were obtained by dividing the emission
$I_c(\theta_e)$ by the total emission spectrum $I_{tot}$ and
correcting for the angular aperture of the collecting lens
$\Omega_L$. For the frequency $\omega$ = 15000 cm$^{-1}$, below
the stop band, the distribution follows the Lambertian
distribution and is similar to the exit distribution from the
nonphotonic sample (Fig.~\ref{fig:det-eff}). For the frequencies
above the red edge of the stop band we observe strongly
non-Lambertian behaviour. For the frequency $\omega$ = 16500
cm$^{-1}$ emission is suppressed relative to the Lambertian
distribution in the range of the exit angles from $\theta_e = 0^0$
to $20^0$. This range moves to larger exit angles for the
frequency $\omega$ = 17400 cm$^{-1}$ in qualitative agreement with
Bragg's law. For $\omega$ = 18350 cm$^{-1}$ the suppression
observed around $\theta_e = 40^0$ is preceded by a considerable
increase of emission in the angle range $0^0$ to $20^0$. A
qualitative explanation of this effect is as follows. Some escape
directions are blocked by internal Bragg diffraction, and
diffusion eventually distributes this back-reflected light along
the remaining directions. Thus, light is more likely to exit the
crystal along these allowed directions. From
Figures~\ref{fig:spectra-ratios} and~\ref{fig:escape-angular} we
conclude that the escape function is in excellent agreement with
the measured angle-dependent spectra. To the best of our
knowledge, the current work provides the first quantitative
modelling of spontaneous emission spectra in 3D photonic crystals.

\subsection{Spontaneous emission from quantum dots in titania inverse opals}
\begin{figure}
\includegraphics[width=0.35\textwidth]{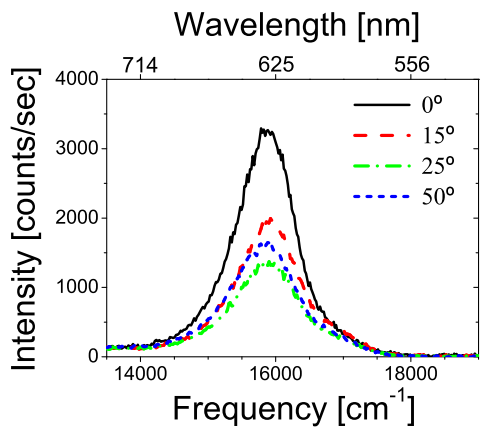}
\caption{(color online) \label{fig:q-dots-spectra} Emission
spectra of CdSe quantum dots in a titania inverse opal with a
lattice parameter $a$ = $500$ nm. The black curve is obtained at
$\theta_e = 0^0$, the red dashed curve at $\theta_e = 15^0$, the
green dash-dotted curve at $\theta_e = 25^0$, and the blue
short-dashed curve at $\theta_e = 50^0$.}
\end{figure}

Titania inverse opals possess a larger relative width of the
first-order stop band ($\Delta\omega/\omega \approx 0.16$) than the
polystyrene opals owing to their inverse structure and high
refractive index contrast ($m = 2.7$). The concomitant large
modifications of the LDOS makes the inverse opals very attractive for
control of propagation and spontaneous emission of
light~\cite{Koenderink02,Lodahl04}. Figure~\ref{fig:q-dots-spectra}
shows emission spectra of CdSe quantum dots in a titania inverse opal
with lattice parameter $a = 500$ nm for selected detection angles
$\theta_e$. No significant changes in the spectral shapes due to
internal Bragg diffraction are observed, because the relative
spectral width of the light sources ($\Delta\omega/\omega < 0.06$) is
considerably smaller than the width of the stop band of the photonic
crystal. This shows that the escape distribution $P(\omega,\theta_e)$
does not vary significantly within the frequency range of the quantum
dot spectrum. In contrast, a strong angular dependence of the
emission intensity is apparent in Fig.~\ref{fig:q-dots-spectra}. As a
consequence, effects of Bragg diffraction are most convincingly
observed by recording the angular dependencies at the spectral maxima
of the emission spectra.

\begin{figure*}
\includegraphics[width=0.75\textwidth]{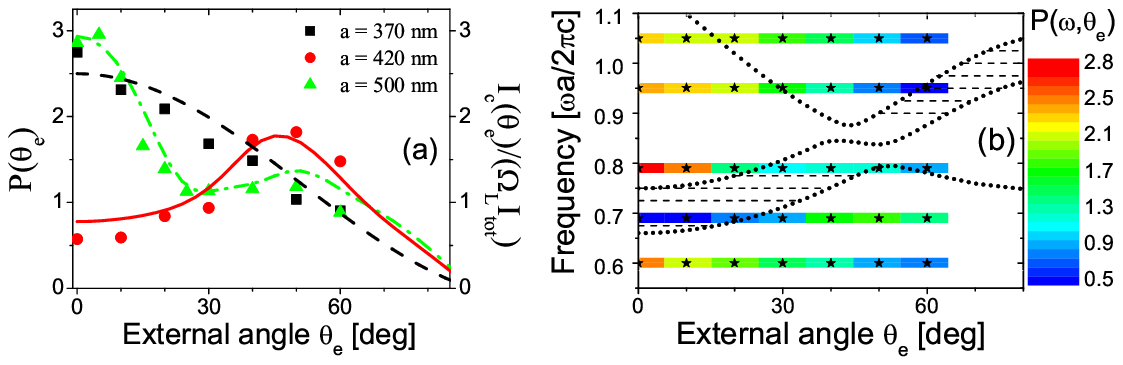}
\caption{(color) \label{fig:titania-angular-bands} (a) Escape
distribution as a function of the exit angle $\theta_e$ for
titania inverse opals with lattice parameters \emph{a} = 370 and
420 nm at $\omega$ = 16390 cm$^{-1}$ ($\lambda$ = 610 nm, black
squares and red circles, respectively), and with a lattice
parameter \emph{a} = 500 nm at $\omega$ = 15870 cm$^{-1}$
($\lambda$ = 630 nm, green triangles). The corresponding curves
represent the calculated distributions. (b) Photonic band
structure for the inverse opals (black dotted curves). The hatched
regions indicate the stop band caused by Bragg diffraction by
(111) lattice planes. The horizontal bars represent the reduced
center frequencies of the quantum-dot emission from the crystals
with lattice parameters (bottom to top) \emph{a} = 370, 420, 500,
580, and 650 nm. The colours of the bars indicate the values of
the escape function $P(\omega,\theta_e)$ obtained from the
measurements at exit angles shown by the black stars.}
\end{figure*}

In Figure~\ref{fig:titania-angular-bands}(a) we present escape
distributions from titania inverse opals with lattice parameters $a =
370$ and 420 nm at frequency $\omega = 16390$ cm$^{-1}$, and with a
lattice parameter $a = 500$ nm at $\omega = 15870$ cm$^{-1}$. Both
measured (symbols) and calculated (curves) values are shown. For the
crystal with the lattice parameter $a = 370$ nm, for which the center
frequency of the quantum dot spectrum lies below the stop band, the
escape function follows the Lambertian distribution. A large
deviation from the Lambertian distribution is observed for the
quantum dot emission from the crystals with the other two lattice
parameters. In the crystals with $a = 420$ nm, the emission is
strongly reduced in the range of the angles from $\theta_e = 0^0$ to
$35^0$, and it is enhanced at higher exit angles. For the crystals
with $a = 500$ nm, the suppression is shifted to the range of
$\theta_e = 20^0$ to $45^0$, as expected for photonic gaps at higher
reduced frequency ($a/\lambda=\omega a/2\pi c$), and in excellent
agreement with our theoretical predictions. The stop-band ranges are
noticeably wider than that in the case of the polystyrene opals
(Fig.~\ref{fig:escape-angular}), which is due to a wider frequency
range of the stop band in the titania inverse opals. To the best of
our knowledge, this is the first demonstration of photonic crystal
bands in the emission spectra of confined excitons in quantum dots.


Figure~\ref{fig:titania-angular-bands}(b) shows the photonic band
structure for a titania inverse opal. The hatched regions indicate
the stop band caused by Bragg diffraction by (111) lattice planes. In
the angular range from $\theta_e = 35^0$ to 55$^0$, multiple Bragg
wave coupling from (111) and (200) diffracted waves takes
place~\cite{vDriel00,Schriemer01}. The horizontal bars represent the
reduced center frequencies of the quantum dot emission from the
inverse opals with the lattice parameters \emph{a} = 370, 420, 500,
580, and 650 nm. The colours of the bars indicate the measured values
of the escape function $P(\omega,\theta_e)$. For reduced frequencies
around the stop bands, it is seen that inhibited escape probability
appears in the angular ranges of the stop bands, whereas enhanced
escape is found outside the stop bands. Hence, the photonic crystals
are seen to `funnel' light along certain allowed directions.

For experiments on quantum dot emission in photonic crystals, the
good agreement between the experimentally obtained escape
distributions and the calculated ones confirms that the light
emanating from inside the crystals is diffuse. It also confirms that
the observed emission is dominated by sources inside the bulk of the
crystal. We can exclude that light sources on the sample surface
contribute significantly: their emission would give rise to an
angle-independent component of the intensity that is not observed.
Furthermore, observation of stop bands in emission spectra is
important for successful lifetime experiments or other
quantum-optical studies of light sources in photonic crystals. The
stop bands are evidence that the emission from the light sources is
strongly coupled to the photonic crystals, and are a prerequisite for
time-resolved experiments of changes in the emission decay rate
caused by a modified LDOS \cite{Lodahl04}.

\begin{figure}
\includegraphics[width=0.35\textwidth]{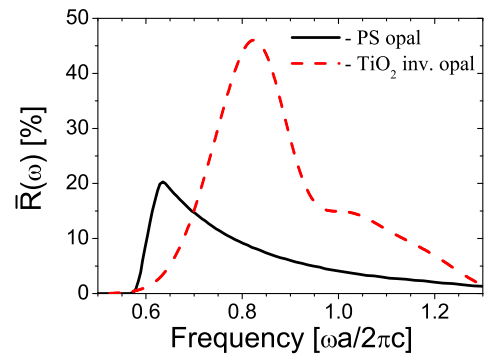}
\caption{(color online) \label{fig:meanR-PS-titania} The
angle-averaged internal reflectivity $\bar{R}(\omega)$ for the
polystyrene opals (black curve) and for the titania inverse opals
(red dashed curve) according to the diffusion model, in which only
the first-order Bragg diffraction is taken into account.
$\bar{R}(\omega)$ determines the enhancement of the escape
probability outside a stop-band direction.}
\end{figure}

Based on the close accordance of the experiments with the model, we
can extract the angle-averaged internal-reflection coefficient
$\bar{R}(\omega)$. Figure~\ref{fig:meanR-PS-titania} shows that
$\bar{R}(\omega)$ is as large as $50\%$ for the titania inverse opals
and up to $20\%$ for the polystyrene opals. The internal-reflection
coefficient varies strongly with frequency in contrast to the
frequency-independent $\bar{R}$ in random media. The coefficient
increases with solid angle for Bragg reflection, starting from the
low-frequency edge of the L-gap. For the opals, the maximum $\bar{R}$
occurs at the high-frequency edge of the L-gap, where the reflecting
stop bands extend over the largest solid angle~\cite{Thijssen99}. For
the inverse opals, the maximum $\bar{R}$ occurs at higher reduced
frequencies in the range of multiple Bragg wave coupling ($\omega
a/2\pi c \sim 0.85$)~\cite{Schriemer01}. The shoulder near $1.0$ is
attributed to the inclusion of (200) reflection condition in our
model. In a more elaborate escape-model with additional diffraction
conditions, we may expect additional peaks in the angle-averaged
reflection coefficient at even higher frequencies. Since the inverse
opals interact stronger with light than the opals, their stop bands
are wider and hence the angle averaged reflectivity is larger, in
agreement with our observations.

A closer consideration of the reflectivity coefficients can serve to
optimize the spontaneous emission yield of light sources (atoms, dyes
or quantum dots) embedded inside thick photonic crystals ($L > l$).
Such an optimization can be achieved either via the excitation of the
sources, via their emission, or both. First, the excitation
efficiency can be increased by realizing that increased escape
probability also implies an increased probability for excitation
light to enter a photonic crystal. Thus, by tuning an excitation beam
to frequencies and angles of high escape, the combined action of
diffusion and Bragg diffraction retain relatively more excitation
light inside the sample, thus increasing the probability for
spontaneous emission of the embedded light sources. Secondly,
spontaneously emitted radiation is efficiently channelled out of the
sample along particular directions. This occurs when the lattice
parameter of the photonic crystal is chosen such that the emission
frequencies are in the range of enhanced $\bar{R}(\omega)$. A clear
example of enhanced escape is apparent in
Fig.~\ref{fig:spectra-ratios} at $\theta_e = 0^0$ near 18000
cm$^{-1}$. In the ultimate case of a photonic band gap, it has even
been predicted that the diffuse emission is extremely directional,
see Ref.~\cite{Koenderink03}. Thirdly, one can envision situations
where \emph{both} excitation and emission are enhanced: In
Fig.~\ref{fig:titania-angular-bands}(b) enhanced escape probability
occurs both at $\omega a/2\pi c \sim 0.8$ and $\theta_e = 0^0$ and at
$\omega a/2\pi c \sim 0.7$ and $\theta_e = 50^0$. Thus, by tuning the
excitation to the former condition and the emission to the latter,
the spontaneous emission yield is expected to be enhanced by at least
a factor of two. Further improvements should be feasible in photonic
crystals with even larger $\bar{R}$.

\section{Conclusions}
We have presented experimental data on angular resolved emission from
light sources embedded in efficient 3D photonic crystals. The
experiments were compared in detail to a recently developed model of
light transport in real photonic crystals that are influenced by
disorder. Our model is based on diffusion of light due to scattering
(disorder) combined with angle- and frequency-dependent internal
reflections (order). Good quantitative agreement between experiment
and theory confirms that the details of the emission spectra are
determined by the intricate interplay of order and disorder.
Properties of the stop bands, such as their frequency range,
magnitude, and angular dependence, are extracted from the experiment
by analyzing the emission escape function. The \emph{enhanced} escape
probability for emission along directions outside the stop bands is
explained by the angle-averaged internal-reflection coefficient
$\bar{R}(\omega)$. The diffuse and angular-dependent nature of light
escaping from the photonic crystals proves that the light comes from
emitters \emph{inside} the crystals. By measuring the escape
functions of the quantum dot emission from the titania inverse opals,
we have for the first time revealed clear stop bands in the quantum
dot emission spectra, confirming that the confined excitons
experience optical confinement. The quantitative agreement between
experiment and theory demonstrates that light propagation and
spontaneous emission in real 3D photonic crystals is well understood.

\section{Acknowledgments}
We thank Floris van Driel and Dani\"{e}l Vanmaekelbergh (University
of Utrecht) for preparation of the quantum dots, L\'{e}on Woldering
for photonic crystal preparation, Arie Irman and Karin Overgaag for
experimental assistance, and last but not least Femius Koenderink for
band-structure calculations and discussions on escape functions. This
work is a part of the research program of the "Stichting voor
Fundamenteel Onderzoek der Materie (FOM)", which is financially
supported by the "Nederlandse Organisatie voor Wetenschappelijk
Onderzoek (NWO)".

\begin{figure}
\includegraphics[width=0.35\textwidth]{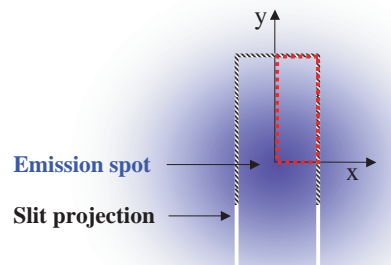}
\caption{(color online) \label{fig:det-eff2} Real-space cartoon of
the projection of the spectrometer slit (striped rectangle) on the
sample surface overlapped with the emission spot (in blue). The
quarter of the slit projection (red dashed rectangle) has one of
the corners in the emission-spot center. The width of the
projection (along \emph{x}) increases at larger detection angles
$\theta_e$.}
\end{figure}
\section{Appendix: Model of detection efficiency of emission set-up}
The aim of this appendix is to explain the difference between the
calculated escape distribution and the measured intensity
illustrated in Fig.~\ref{fig:det-eff}. This difference appears
because the width of the projection of the collection optics (the
spectrometer slit) on the sample increases with the angle
$\theta_e$. The sample surface is placed in the focus of the
collecting lens $L_1$, the spectrometer slit is in the focus of
the imaging lens $L_2$ (\emph{cf.} Fig.~\ref{fig:set-up}). The
only emission collected emanates from the region confined by the
slit projection on the sample surface, see
Fig.~\ref{fig:det-eff2}. This means that the spectrometer collects
light from a larger region on the surface at larger detection
angles, and that the measured angle-dependent intensity should be
corrected for the detection efficiency of the set-up. The
detection efficiency is modelled as a ratio $D(\mu_e)$ of the
intensity $B(\mu_e)$ collected from the surface region within the
slit projection (Fig. \ref{fig:det-eff2}) at detection angle
$\theta_e = \cos^{-1}(\mu_e)$ to the intensity $B(\mu_e=1)$
collected at normal angle:
\begin{equation}
D(\mu_e)=\frac{B(\mu_e)}{B(\mu_e=1)}\,,~~~~B(\mu_e)=\int^{x(\mu_e)}_0{\int^{y_0}_0{I(x,y)dxdy}.}
\end{equation}
We take into account that the integration runs over a quarter of
the slit, as $x(\mu_e)=x_0/\mu_e$ is the half-width of the slit
projection, $x_0$ and $y_0$ are the half-width and the half-height
of the slit projection at $\mu_e=cos(\theta_e)=1$, respectively.
Typical values of $x_0$ and $y_0$ in the experiments are 50 $\mu
m$ and 1 mm. It is assumed that the intensity of diffuse light on
the sample surface around the pump beam varies as $I(r)\propto
l^2/(l^2+r^2)$, where $r^2=x^2+y^2$ is the distance from the axis
of the pump beam along the sample surface, and $l$ is the mean
free path of light in the sample.


\begin{thebibliography}{26}
\bibitem{bleaching}{Bleaching is a photoinduced damage of dye molecules (fluorophores).
Typically, it happens because of oxidation of the fluorophore that
damages the molecule permanently.}
\bibitem{average refr index}
The average refractive index of a photonic crystal is estimated
as: $n_{av} = \varphi \cdot n_{m} + (1 - \varphi)\cdot n_{air}$,
where $\varphi$ is the volume fraction of the solid material,
$n_{m}$ and $n_{air}$ are refractive indices of the solid material
and air, respectively.
\end{thebibliography}
\end{document}